# Frequency ratios of Sr, Yb and Hg based optical lattice clocks and their applications

Masao Takamoto[a,b,c], Ichiro Ushijima[a,b,c], Manoj Das[a,b,c], Nils Nemitz[a], Takuya Ohkubo[a,c,d], Kazuhiro Yamanaka[a,c,d], Noriaki Ohmae[a,c,d], Tetsushi Takano[c,d], Tomoya Akatsuka[a,b,c], Atsushi Yamaguchi[a,b,c], Hidetoshi Katori[a,b,c,d,*]

[a]*Quantum Metrology Laboratory, RIKEN, Wako, Saitama 351-0198, Japan.*
[b]*RIKEN Center for Advanced Photonics, Wako, Saitama 351-0198, Japan.*
[c]*Innovative Space-Time Project, ERATO, Japan Science and Technology Agency, Bunkyo-ku, Tokyo 113-8656, Japan.*
[d]*Department of Applied Physics, Graduate School of Engineering, The University of Tokyo, Bunkyo-ku, Tokyo 113-8656, Japan.*

**Abstract**

This article describes the recent progress of optical lattice clocks with neutral strontium ($^{87}$Sr), ytterbium ($^{171}$Yb) and mercury ($^{199}$Hg) atoms. In particular, we present frequency comparison between the clocks locally via an optical frequency comb and between two Sr clocks at remote sites using a phase-stabilized fibre link. We first review cryogenic Sr optical lattice clocks that reduce the room-temperature blackbody radiation shift by two orders of magnitude and serve as a reference in the following clock comparisons. Similar physical properties of Sr and Yb atoms, such as transition wavelengths and vapour pressure, have allowed our development of a compatible clock for both species. A cryogenic Yb clock is evaluated by referencing a Sr clock. We also report on a Hg clock, which shows one order of magnitude less sensitivity to blackbody radiation, while its large nuclear charge makes the clock sensitive to the variation of fine-structure constant. Connecting all three types of clocks by an optical frequency comb, the ratios of the clock frequencies are determined with uncertainties smaller than possible through absolute frequency measurements. Finally, we describe a synchronous frequency comparison between two Sr-based remote clocks over a distance of 15 km between RIKEN and the University of Tokyo, as a step towards relativistic geodesy.
*Keywords:* Time and frequency metrology; atomic clock; optical lattice clock; constancy of fundamental constants; relativistic geodesy

## 1. Introduction

Optical lattice clocks [1] benefit from a low intrinsic quantum-projection noise (QPN) limit [2] due to the simultaneous interrogation of a large number of atoms, which are trapped in an optical lattice operated at the "magic wavelength" to largely cancel out light shift perturbation in the clock transition [3]. Taking full advantage of the clock stability either with interrogation lasers stabilized to ultra-stable optical resonators [4] or by rejecting the laser noise through the application of synchronous interrogation schemes [5], optical lattice clocks have reached instabilities on the level of $10^{-16}$ at one second [6-8]. Such stabilities allow achieving $10^{-18}$ uncertainty in a few hours of clock operation, which expedites investigation of systematic uncertainties, such as collisional shift [6], multipolar and higher order light shifts in the optical lattice [9], and the blackbody radiation shift [10]. These evaluations have allowed optical lattice clocks to reach inaccuracies approaching $10^{-18}$ [10, 11]. Although the primary caesium standards still continue their development and now state systematic uncertainties as low as $1.1 \times 10^{-16}$ [12], it is now the uncertainty of the SI second itself, that restricts the measurement of the absolute frequencies of optical standards [13, 14]. Direct optical comparisons [6-11, 13-20] are the only way for optical clocks to investigate their superb performance beyond the SI limit.

Comparison of two clocks referring to the same atomic transition gives null-measurements, which facilitate investigation of systematic uncertainties as well as the clocks' stability [10, 11, 16]. When measuring the same clocks between two distant laboratories connected through fibre links [21, 22], general-relativistic frequency shifts make such clocks a probe for the gravitational potential [23], which promise new applications such as relativistic geodesy [24, 25]. On the other hand, comparisons of optical clocks based on different atomic elements have been performed by utilizing optical frequency-comb technique [26, 27], which may determine a frequency ratio [15] with higher precision than absolute frequencies based on the SI second. Such a frequency ratio can be shared with full accuracy and tested in any laboratory across the world. To collect the measured frequency ratios, the Consultative Committee for Time and Frequency (CCTF) of the International Committee for Weights and Measures (CIPM) has proposed the creation of a "frequency matrix" [28] to complement the list of secondary representations of the second that has been maintained since 2006. Such a matrix would allow the calculation of optimal synthetic values for frequency ratios that have not yet been measured directly, as well as improved consistency checks. These efforts should be an essential step towards a redefinition of the SI second. In addition, a collection of frequency ratios over time will provide an invaluable resource in the search for a temporal variation of the fundamental constants [15, 19, 29, 30].

* Corresponding author. Tel.: +81-48-462-1966; fax: +81-48-467-1965.
*E-mail address:* hkatori@riken.jp (H. Katori).

In this article, we describe recent progress of optical lattice clocks with neutral strontium (Sr), ytterbium (Yb) and mercury (Hg) atoms and present frequency comparisons between them. We also describe remote comparisons between Sr clocks located at RIKEN and the University of Tokyo over a 30-km-long phase-stabilized fibre link. In particular, we emphasize that synchronous comparison between optical lattice clocks [1, 8] allows accessing reduced uncertainties in shorter averaging time, facilitating the gain of useful information from frequency comparisons.

## 2. Cryogenic strontium optical lattice clocks as a frequency anchor with $7 \times 10^{-18}$ uncertainty

One of the most serious obstacles for optical lattice clocks with $^{87}\mathrm{Sr}$ to attain an uncertainty at low $10^{-18}$ has been the ac Stark shift due to the blackbody radiation (BBR). This introduces a fractional frequency shift of a few times $10^{-15}$ in a room temperature environment, which is orders of magnitude larger than that of $\mathrm{Al}^{+}$ ion-clocks [31]. The BBR shift for the $^1S_0 - {}^3P_0$ clock transition of Sr was calculated [3, 32, 33], and experimentally evaluated to be $-2.2778(23)$ Hz at $T = 300$ K [34]. Similar evaluations have also been performed for Yb atoms [35, 36]. Such investigations should allow correcting the room-temperature BBR shift within uncertainties of $5 \times 10^{-18}$ for Sr [33, 34] and $1 \times 10^{-18}$ for Yb [37]. However, accurate determination of the ambient temperature is an experimental challenge, as the temperature needs to be measured and controlled to within $\Delta T = 14$ mK for Sr and $\Delta T = 30$ mK for Yb to achieve clock uncertainty of $1 \times 10^{-18}$.

We therefore take a straightforward approach of interrogating the atoms inside a cryogenic environment as assumed in Ref. [3]. As the energy density $\langle E^2\rangle_{\mathrm{BBR}}$ of BBR varies as $T^4$ following the Stefan-Boltzmann law, the BBR shift $\nu_{\mathrm{BBR}} \approx -(1/2)\Delta\alpha\langle E^2\rangle_{\mathrm{BBR}}$ reduces as $T^4$ for the surrounding temperature $T$ [38], where $\Delta\alpha$ is the difference of dipole polarizability in the clock transition. Consequently, temperature sensitivity of the BBR shift reduces as $T^3$, which significantly relaxes the accuracy of temperature measurement for lower temperature. For Sr atoms, when operating at $T = 95$ K, with a corresponding BBR shift correction of $5 \times 10^{-17}$, the fractional uncertainty due to the BBR shift can be reduced to $1 \times 10^{-18}$ by controlling the temperature to within $\Delta T = 0.5$ K.

The cryogenic environment is realized by surrounding the region for clock spectroscopy with a cold chamber of volume $\sim 6\ \mathrm{cm}^3$ [Fig. 1(a)]. The cryo-chamber is cooled down to 95 K by a Stirling refrigerator, which is actively temperature-stabilized to within a few mK. Ultracold $^{87}\mathrm{Sr}$ atoms [39], which are loaded into a one-dimensional lattice at the magic wavelength of $\lambda_{\mathrm{m(Sr)}} = 813.4$ nm outside the cryo-chamber, are transported over 23 mm into the middle of the chamber by a moving lattice. The chamber has two apertures with diameters of $\phi_1 = 0.5$ mm and $\phi_2 = 1$ mm to introduce atoms and lasers. Inside the chamber, atoms are spin-polarized and subsequently excited by a Rabi pulse on the $^1S_0 - {}^3P_0$ clock transition at 698 nm. Then, the atoms are transported back to the loading position to measure the excitation probability of the clock transition, which is used to steer the clock laser frequency.

Two cryo-clock setups, Sr-1 and Sr-2, are equipped with temperature-controlled chambers kept at $T_1$ and $T_2$. From the frequency difference $\Delta\nu(T_1, T_2) = \nu_2(T_2) - \nu_1(T_1)$ of these two clocks, we have evaluated the BBR shift. In order to measure the frequency difference of two clocks precisely in a short averaging time, we operate the clocks synchronously by interrogating the clock transitions with probe pulses from a common clock laser during the same time period. Such synchronous operation of two clocks allows common-mode rejection of clock laser phase noise and facilitates the evaluation of the clocks' frequency difference, as demonstrated in Ref. [8]. The inset in Fig. 1(c) shows the spectra of the clock transitions obtained by scanning the clock laser frequencies simultaneously for Sr-1 and Sr-2. The blue circles show the spectrum of Sr-1 at $T_1 = 95$ K and the orange triangles show that for Sr-2 at $T_2 = 296$ K, which resolve the room-temperature BBR shift of 2 Hz. Red circles in Fig. 1(c) shows the temperature dependence of the BBR shift $\Delta\nu(95\ \mathrm{K}, T_2)$ by varying the temperature of Sr-2 from $T_2 = 95$ K to 296.2 K, where the temperature of Sr-1 is fixed at $T_1 = 95$ K as a reference. The temperature dependence of the BBR shift is given by

$$\nu_{\mathrm{BBR}}(T) = \nu_{\mathrm{stat}}(T/T_0)^4 + \nu_{\mathrm{dyn}}(T/T_0)^6 + O(T/T_0)^8,$$

where $\nu_{\mathrm{stat}}$ and $\nu_{\mathrm{dyn}}$ are the static and dynamic contributions at $T_0 = 300$ K [32]. The data points follow the temperature dependence of $\Delta\nu(95\ \mathrm{K}, T) = \nu_{\mathrm{BBR}}(T) - \nu_{\mathrm{BBR}}(95\ \mathrm{K})$. The red line is the fitted curve to the data points with $\nu_{\mathrm{dyn}}$ as the fitting parameter. Applying $\nu_{\mathrm{stat}} = -2.13023$ Hz taken from the measured differential static-polarizability $\Delta\alpha$ of the clock transition [34], the best fitting is obtained for $\nu_{\mathrm{dyn}} = -0.1480(26)$ Hz [10], which agrees with the values derived by theoretical calculations in Ref. [33, 34] within $1\sigma$ uncertainty.

Operation of both the clocks at 95 K allows us to evaluate the cryo-clocks' systematic uncertainties. The Allan deviation of the frequency difference of two clocks in a synchronous operation reduces with the integration time $\tau$ as $\sigma_y(\tau) = 1.8 \times 10^{-16}(\tau/s)^{-1/2}$ and reaches $2 \times 10^{-18}$ for $\tau = 6000$ s, as shown in Fig. 3. The obtained Allan deviation is about 1.6 times the QPN limit for 300-ms-long interrogation time and $N = 10^3$ atoms. With the systematic evaluations as summarized in Table 1, we have achieved a fractional systematic uncertainty of $7.2 \times 10^{-18}$ for a single independent clock. After averaging 11 separate measurements over a month (Fig. 1(d)), the two clocks have reached an agreement to $\Delta\nu/\nu_0 = \left(-1.1 \pm 2.0(\mathrm{stat}) \pm 4.4(\mathrm{sys})\right) \times 10^{-18}$ [10], after including systematic corrections.

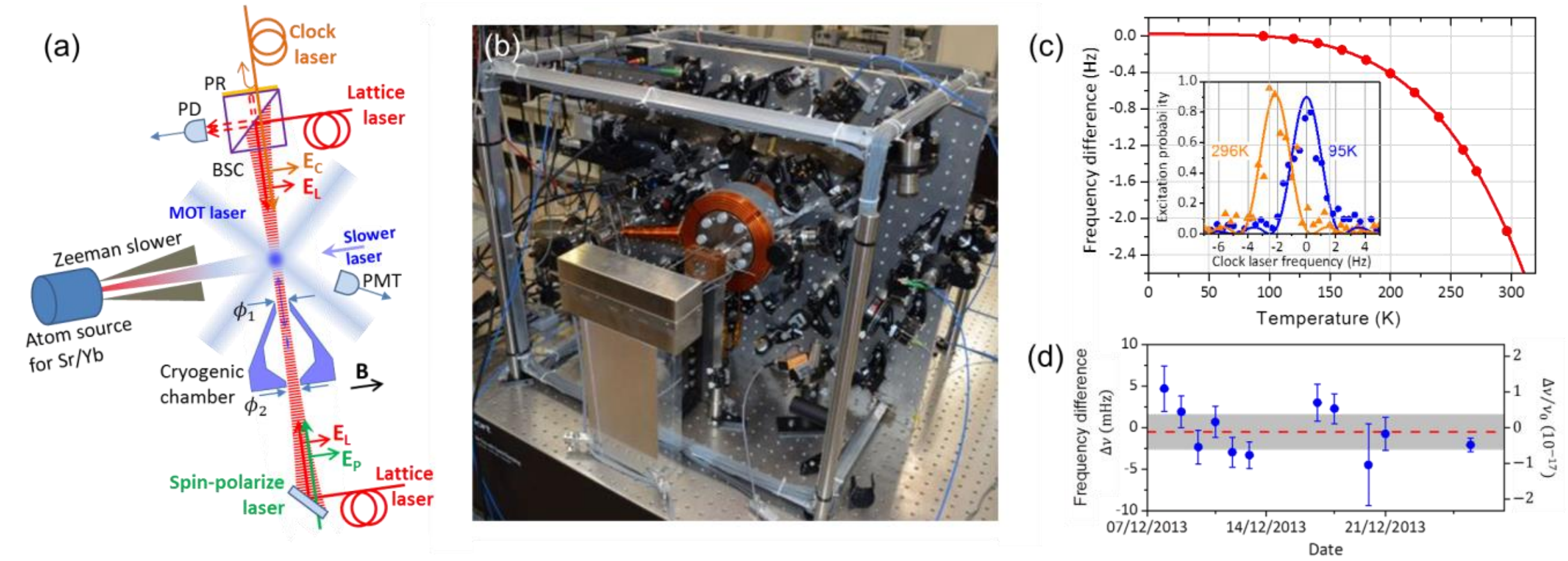

Fig. 1. (a) A schematic for a cryogenic optical lattice clock. A moving lattice transfers the atoms into the cryogenic chamber through the aperture. Partial reflector (PR) attached on a beam-splitter cube (BSC) works as a reference for the Doppler cancellations for the clock and lattice lasers. The system is compatibly-designed for Sr and Yb optical lattice clocks and the atom source for Sr-2 includes both Sr and Yb. $\mathbf{E}_{\mathrm{c}}$, $\mathbf{E}_{\mathrm{L}}$, and $\mathbf{E}_{\mathrm{p}}$ show the electric field vectors for clock, lattice and pumping lasers. PD, photo diode; PMT, photomultiplier tube. (b) A photograph of a cryogenic optical lattice clock. (c) Temperature dependence of the BBR shift for Sr. The inset shows the clock spectra observed at $T_1 = 95$ K (circles) and $T_2 = 296$ K (triangles). The frequency difference $\Delta\nu(95\,\mathrm{K}, T_2)$ is measured by varying the temperature $T_2$, while keeping the temperature $T_1 = 95$ K. The line shows the fit to the data points to obtain a dynamic contribution of $\nu_{\mathrm{dyn}} = -0.1480(26)$ Hz. (d) The averaged frequency difference between the two Sr cryo-clocks. Error bars represent the $1\sigma$ statistical uncertainty for each measurement. Dashed line and shaded region represents the averaged frequency and the total uncertainty of the measurements, respectively.

## 3. Frequency comparisons among optical lattice clocks based on different atomic elements

### 3.1 Overview of the local frequency link using an Er-fibre based frequency comb

A frequency comb provides convenient access for comparing clock transitions at 266 nm (Hg), 578 nm (Yb) and 698 nm (Sr) as described here. We use a multi-branch Er-doped fibre comb developed at NMIJ [40, 41], which features a high feedback control bandwidth of 1 MHz by an intra-cavity electro-optic modulator to achieve stability better than $1 \times 10^{-15}$ at 1 s. Figure 2 shows schematics for the clock comparisons.

An external cavity diode laser (ECDL-1) at 698 nm, locked to a 40-cm-long cavity with a spacer made of Asahi Glass AZ, serves as a master laser for the system. An ECDL-2 at 1397 nm, which is the sub-harmonic wavelength of the master laser, is frequency-doubled and phase-locked to the master laser. A beat note between ECDL-2 and the nearest comb line is used to stabilize the repetition rate $f_{\mathrm{rep}}$ of the comb. By stabilizing the carrier-envelope offset frequency $f_{\mathrm{CEO}}$ using the self-referencing technique [42], we fix frequencies of all comb lines as $\nu_n = f_{\mathrm{CEO}} + n f_{\mathrm{rep}}$ for $n$-th longitudinal mode. Any frequency in the spectral range from 1000 nm to 2100 nm can be determined by measuring the beat signal with the comb line $\nu_n$, which is related to the Sr clock frequency $\nu_{\mathrm{Sr}}$ via the ECDL-1 and 2.

Clock lasers for Yb and Hg clocks are derived from an ECDL-3 at 1156 nm and a fibre laser at 1063 nm respectively, whose wavelengths are conveniently phase-locked to the Er-comb. This stabilization transfers the stability of the master laser to Yb and Hg clock lasers [43], allowing synchronous operation to partially cancel out the laser frequency noise in frequency ratio measurements [8]. Frequency shifters inserted in each clock laser allow independent control of their frequencies to be stabilized to the clock transitions. These frequency offsets contribute less than $10^{-6}$ to the total optical frequencies. To allow ratio measurements with instabilities in the $10^{-18}$ range, we use a GPS-disciplined BVA-quartz oscillator with an instability below $10^{-11}$ to provide a reference signal for the DDSs (direct digital synthesizers) that drive the frequency shifters.

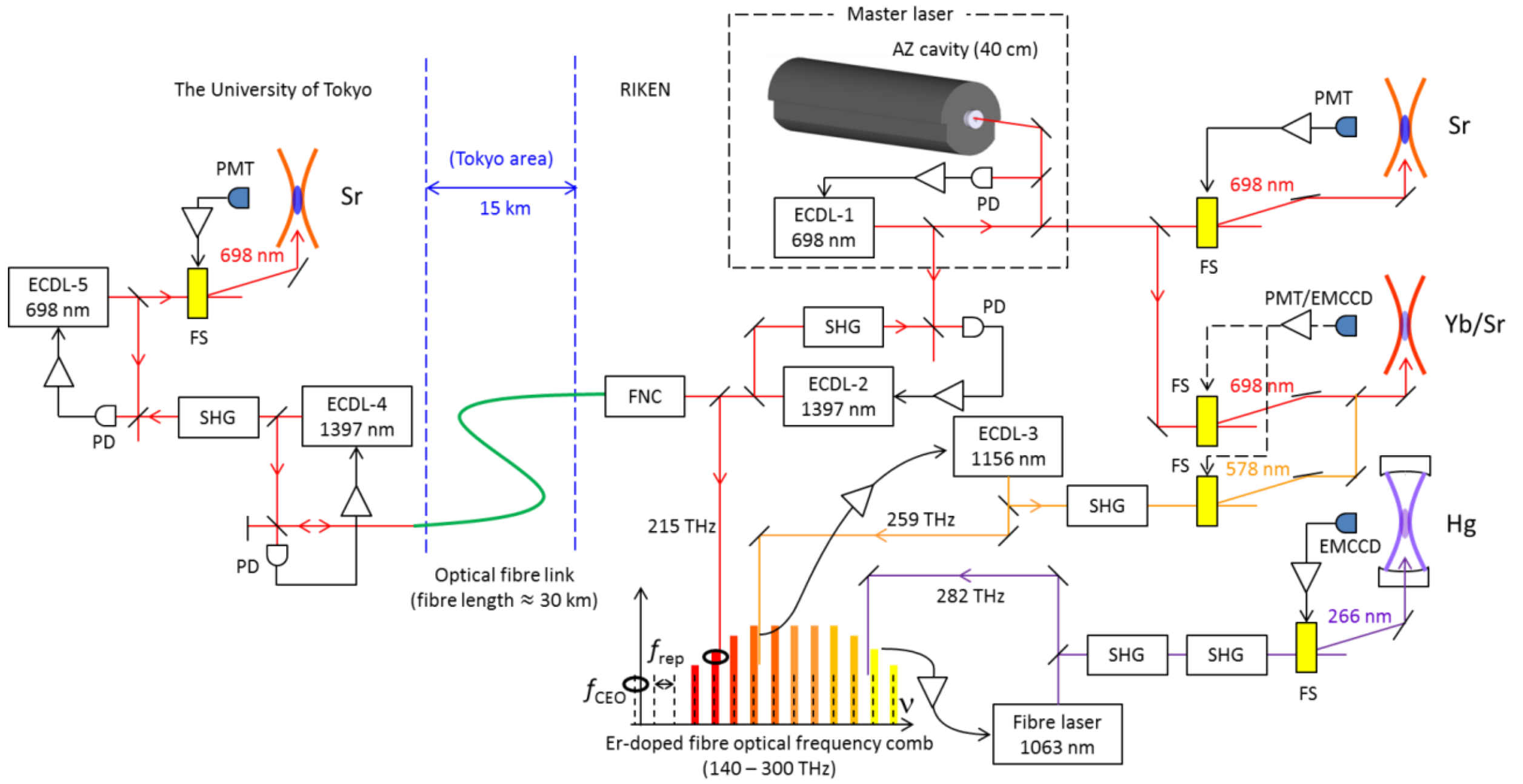

Fig. 2. A schematic for frequency comparison of Hg, Yb, and Sr clocks at 266 nm, 578 nm and 698 nm located at RIKEN and the University of Tokyo. By using an Er-doped frequency comb, all clock lasers are referenced to a single master laser locked to a 40-cm-long cavity. ECDL, external-cavity diode laser; SHG, second-harmonic generation; FS, frequency shifter; PD, photo diode; PMT, photomultiplier tube; EMCCD, electron-multiplying CCD camera; FNC, fibre noise canceler.

### 3.2 A cryogenic optical lattice clock with ytterbium

Compared with $^{87}$Sr that has nuclear spin of $I = 9/2$, $^{171}$Yb with $I = 1/2$ significantly reduces the number of Zeeman substates and is free from the tensor light shift. In addition, its sensitivity to BBR is smaller than that of Sr. Optical lattice clocks with Yb atoms are now operated [44-46] or being developed [47, 48] by multiple groups, and a pair of such clocks has recently demonstrated an instability of $1.6 \times 10^{-18}$ [7].

At RIKEN, one of the Sr-clock systems described in Sect. 2 is used to operate a cryogenic Yb-clock. By loading the atomic oven with a mixture of Sr and Yb metals, we selectively laser-cool and trap target atoms (Sr or Yb) by switching between two sets of lasers. This is a straightforward process that does not require any realignments of the chamber-side optics and takes less than two hours. The operating scheme for the $^{171}$Yb-clock is similar to that described in Ref. [44]. Typically $N \approx 1200$ $^{171}$Yb atoms are trapped in a near-vertical optical lattice at $\lambda_{\mathrm{m(Yb)}} = 759.4$ nm with a potential depth of $110\,E_{\mathrm{r}}$, where $E_{\mathrm{r}} = (h/\lambda)^2/(2m)$ is the photon recoil energy for the lattice wavelength $\lambda = \lambda_{\mathrm{m(Yb)}}$ and the atomic mass $m$. Using a moving lattice, the atoms are transported to the centre of the cryogenic chamber, where they are spin-polarized by optical pumping on the $^1S_0 - {}^3P_1$ transition and optionally sideband-cooled to the vibrational ground state by exciting the $^1S_0 - {}^3P_0$ clock transition followed by quenching on the $^3P_0 - {}^3D_1$ transition at 1388 nm. After interrogation by a 160-ms-long Rabi pulse, the atoms are transported back to the loading region for detection. A series of images taken by an EMCCD (electron-multiplying CCD) camera determine ground state and excited state population as well as background counts, in order to deduce the excited state fraction and servo-control the clock laser. Systematic shifts are evaluated by operating the clock in an interleaved mode, where the investigated parameter is changed after every two clock cycles. These measurements show an instability that falls as $2 \times 10^{-15}\,(\tau/\mathrm{s})^{-1/2}$ and reaches $5 \times 10^{-17}$ after $\tau = 2000$ s.

The uncertainty budget for the Yb clock is shown in Table 1. The quadratic Zeeman shift and the probe light shift are determined from the values published in Ref. [44]. The residual BBR shift inside the cryo-chamber and its uncertainty are calculated based on the evaluation of incident radiation performed for the Sr cryo-clock [10] in conjunction with the resulting ac Stark shift analysed in Ref. [36]. The narrow range of accessible lattice depths, limited to a minimum of $70\,E_{\mathrm{r}}$ by excessive loss of atoms and to a maximum of $110\,E_{\mathrm{r}}$ by the available laser intensity, complicates the evaluation of lattice light shifts. Systematic effects resulting from a change in the lattice intensity are magnified by the extrapolation to the full trap depth and the accompanying uncertainties form the dominant contribution to the systematic uncertainty. Our investigation finds a magic frequency at $\nu_{\mathrm{m(Yb)}} = 394\,798\,274(54)$ MHz, where a variation of the lattice intensity does not result in a significant change in the measured clock frequency. The stated light shift correction also accounts for the hyperpolarizability (based on the coefficient determined in Ref. [49]) and for multipolar effects. The latter are investigated using the relation $\Delta\nu_{\mathrm{mp}} \propto (n + 1/2)I^{1/2}$ based on the model presented in Ref. [50]. Sideband cooling allows us to lower the average vibrational quantum number from $\langle n\rangle = 1.3$ to $\langle n\rangle = 0.3$ and set a limit on the magnitude of the effect. To measure the collisional shift, we vary the atom number by changing the duration of the atom loading time into the MOT. At an interrogation time $T_{\mathrm{i}} = 160$ ms, no statistically significant frequency shifts have been observed over repeated measurements at an uncertainty of $3\times10^{-17}$. For typical operating parameters, we estimate the overall systematic uncertainty of the Yb clock to be $\sigma_{\mathrm{sys(Yb)}} = 2.4\times10^{-16}$.

### 3.3 Mercury optical lattice clock

An order of magnitude smaller sensitivity to the BBR compared to Sr and Yb clocks [31] makes $^{199}$Hg (nuclear spin $I = 1/2$) a promising candidate for optical lattice clocks [51-53] in spite of the required wavelengths in UV region which are relatively difficult in terms of generation and handling. Moreover, because of its large nuclear charge $Z_{\mathrm{Hg}} = 80$, the clock transition is relatively sensitive $\sim(\alpha Z)^2$ to the variation of the fine-structure constant $\alpha$ among the potential candidates of optical lattice clocks. Clock frequency measurement of Hg referencing an $\alpha$-insensitive anchor such as Sr may reveal $\Delta\alpha/\alpha = 1.3\,\Delta\nu/\nu_{\mathrm{Hg}}$ [54], where $\Delta\nu/\nu_{\mathrm{Hg}}$ is the fractional change of Hg transition frequency.

Operation of Hg optical lattice clock proceeds as follows: $^{199}$Hg atoms are laser-cooled and trapped on the $^1S_0 - {}^3P_1$ transition at 254 nm [51, 55]. Then, the atoms are loaded into a vertically oriented optical lattice at the magic wavelength of $\lambda_{\mathrm{m(Hg)}} = 362.6$ nm. The optical lattice is formed inside a power build up cavity with an enhancement factor of $\sim 10$ to provide a trap potential depth of $\sim 40\,E_{\mathrm{r}}$. The cavity mirrors are placed outside the vacuum chamber to avoid UV-degradation of the mirror coating occurring in vacuum. We use Brewster windows to provide optical access into vacuum, which allow polarization-selective enhancement of the lattice laser power. In the optical lattice, more than 95 % of atoms are spin-polarized into the clock ground state. The Hg-clock laser at 266 nm is superimposed on the lattice laser. The clock laser's optical path length is fixed by referencing one of the mirrors of the power build-up cavity that determines the trapped atom position to reduce the Doppler noise. We excite the clock transition by a 120-ms-long $\pi$ pulse to obtain a Fourier-limited linewidth of 7.4(4) Hz (FWHM), which corresponds to quality factor $Q \sim 1.5\times10^{14}$. The excited fraction is determined by the laser induced fluorescence on the $^1S_0 - {}^3P_1$ transition and is used to stabilize the clock laser. The overall systematic uncertainty of the Hg clock is evaluated to be $\sigma_{\mathrm{sys(Hg)}} = 7.2\times10^{-17}$ [56].

Table 1. Typical corrections applied for systematic effects and their uncertainties for the Sr, Yb and Hg clocks, given in units of $10^{-18}$ fractional frequency.

| | **$^{87}$Sr** | | **$^{171}$Yb** | | **$^{199}$Hg** | |
|---|---|---|---|---|---|---|
| **Effect** | correction | uncertainty | correction | uncertainty | correction | uncertainty |
| Quadratic Zeeman shift | 117.0 | 0.9 | 69 | 10 | 61 | 9 |
| Blackbody radiation shift | 54.2 | 0.9 | 26 | 1 | 161 | 33 |
| Lattice light shifts | 0 | 4.6 | 185 | 237 | 36 | 61 |
| Probe light shift | 0.047 | 0.023 | -1 | 5 | 0 | < 1 |
| AOM chirp and switching | 0 | < 0.2 | 0 | < 1 | 0 | 10 |
| Servo error | 0 | 3.5 | 0 | 1 | -4 | 3 |
| Density shift | 0.9 | 4.2 | 7 | 15 | 16 | 16 |
| **Systematic total** | **172.3** | **7.2** | **287** | **238** | **269** | **72** |
| | | | **Yb / Sr** | | **Hg / Sr** | |
| Differential grav. red shift | | | 0 | < 1 | -5 | 1 |
| Laser transfer instability | | | 0 | 20 | 0 | 20 |
| **Total uncertainty** | | | | **239** | | **84** |

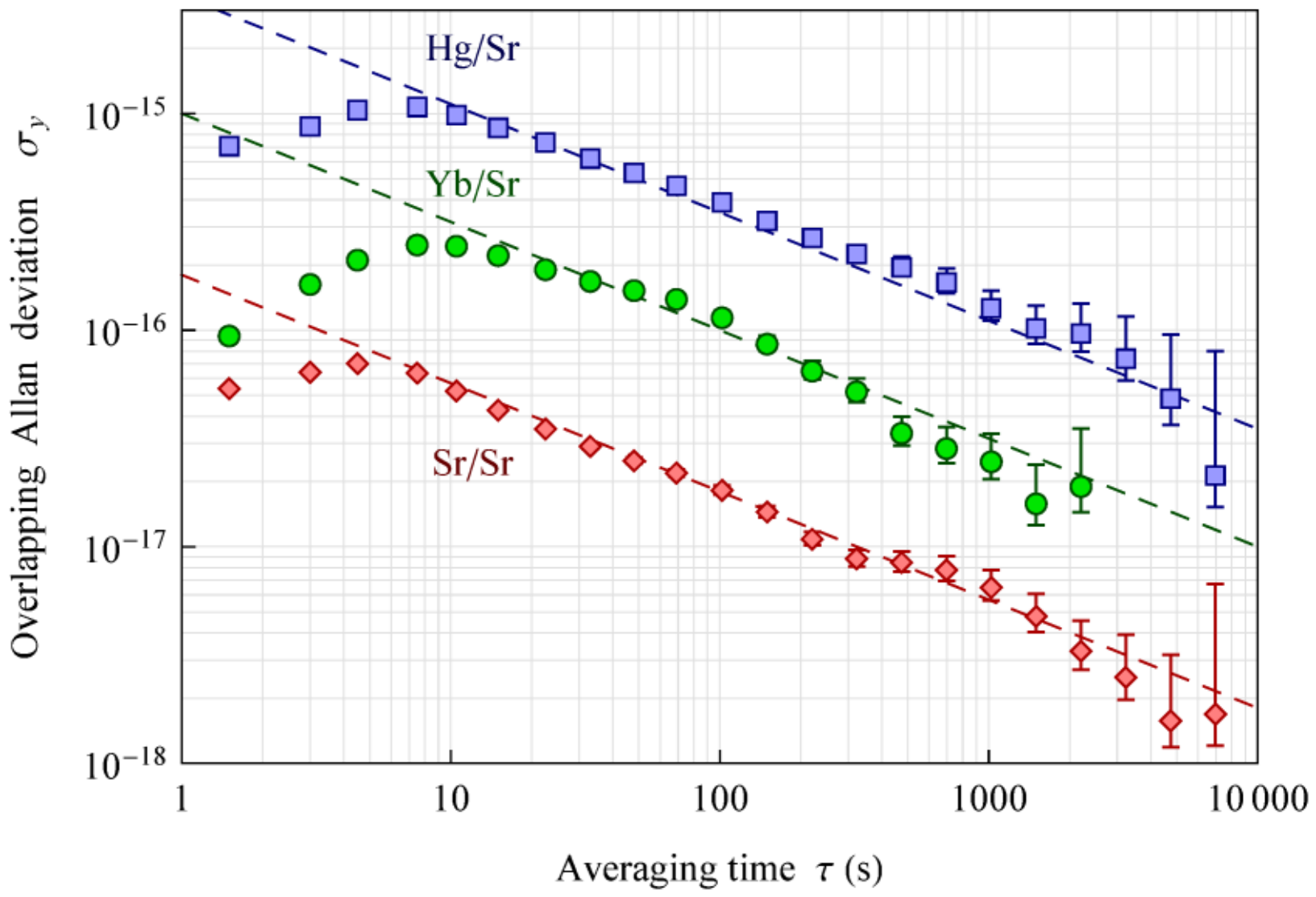

Fig. 3. Instability of clock comparisons. Red diamonds: Taking advantage of the synchronous interrogation technique, the Sr/Sr comparison shows an instability that falls as $\sigma_y(\tau) = 1.8 \times 10^{-16} (\tau/\mathrm{s})^{-1/2}$ with averaging time $\tau$. The Hg/Sr (blue squares) and Yb/Sr (green circles) comparisons are limited by uncorrelated laser noise introduced in imperfect stability transfer and show instabilities falling as $\sigma_y(\tau) \approx 3 \times 10^{-15} (\tau/\mathrm{s})^{-1/2}$ and $\sigma_y(\tau) \approx 1 \times 10^{-15} (\tau/\mathrm{s})^{-1/2}$, respectively.

### 3.4 Frequency ratio measurements of Yb and Hg clock referenced to a Sr clock

We have performed individual measurements of both the Yb/Sr and the Hg/Sr frequency ratios. Figure 3 shows Allan standard deviations for both comparisons with fractional instabilities around $1 \times 10^{-15} (\tau/\mathrm{s})^{-1/2}$ and $3 \times 10^{-15} (\tau/\mathrm{s})^{-1/2}$, respectively. The predicted instability contributions from QPN [2] are $\approx 5 \times 10^{-16} (\tau/\mathrm{s})^{-1/2}$ for the Hg clock and $\approx 2 \times 10^{-16} (\tau/\mathrm{s})^{-1/2}$ for both Yb and Sr clocks when operated at interrogation times of $T_\mathrm{i} = 40$ ms for Hg and $T_\mathrm{i} = 160$ ms for Yb and Sr. Thus, the QPN is not the dominant contribution to the instability in the frequency comparisons of clocks consisted of different atomic elements. The major contributor is the Dick effect [57] that is induced by the periodic interrogation of the clock transitions by lasers with residual phase noise. For the Yb/Sr comparison we partially cancel the noise contribution from the master laser by synchronized application of the interrogation pulses [8]. In order to reduce the additional phase noise induced by the optical path length fluctuations, we apply interferometric phase noise cancellation [58] to many free-space optical paths and to the optical fibres. However, one of the residual phase noise sources is found in the frequency comb setup: Based on measurements of the phase noise between the individual branch amplifiers, we estimate the instability contribution of the frequency comb as $\approx 1 \times 10^{-15} (\tau/\mathrm{s})^{-1/2}$ for typical interrogation cycles. This is consistent with the instability as low as $1 \times 10^{-15} (\tau/\mathrm{s})^{-1/2}$ observed in synchronous measurements of the Yb/Sr frequency ratio. Since the Hg clock is normally operated with an interrogation period of only 40 ms and the clock frequencies differ by a factor of 2.6, synchronous interrogation offers less benefit for a Hg/Sr comparison [8]. For asynchronous operation, we typically observe an instability of $\approx 3 \times 10^{-15} (\tau/\mathrm{s})^{-1/2}$.

For the Yb/Sr frequency ratio, we used a dataset covering 7100 s of measurement time. After correcting for systematic shifts and including a statistical uncertainty of $\sigma_\mathrm{stat} = 1.7 \times 10^{-17}$, we find $\nu_\mathrm{Yb}/\nu_\mathrm{Sr} = 1.207\,507\,039\,343\,337\,76(29)$, corresponding to a fractional uncertainty of $2.4 \times 10^{-16}$. Our measurement is well within the uncertainty of the ratio calculated from the CIPM recommended values [59] as $(\nu_\mathrm{Yb}/\nu_\mathrm{Sr})_\mathrm{CIPM} = 1.207\,507\,039\,343\,339\,9(35)$, but deviates from the ratio of $(\nu_\mathrm{Yb}/\nu_\mathrm{Sr})_\mathrm{NMIJ} = 1.207\,507\,039\,343\,340\,4(18)$ recently measured at NMIJ [45] by 1.5 times their stated uncertainty. Future measurements should either confirm our current result or clarify the source of this deviation.

For the Hg/Sr comparison, the atomic height difference of 5 cm between the Hg and Sr clocks gives a correction of $-5(1) \times 10^{-18}$ to account for the gravitational redshift. The Hg/Sr ratio is given as $\nu_\mathrm{Hg}/\nu_\mathrm{Sr} = 2.629\,314\,209\,898\,909\,60(22)$, corresponding to a fractional uncertainty of $8.4 \times 10^{-17}$, including a statistical uncertainty of $\sigma_\mathrm{stat} = 3.7 \times 10^{-17}$, which is reported in detail in Ref. [56]. Using an earlier absolute frequency measurement performed at LNE-SYRTE [13, 52] , a synthesized ratio of $(\nu_\mathrm{Hg}/\nu_\mathrm{Sr})_\mathrm{SYRTE} = 2.629\,314\,209\,898\,927(15)$ is obtained, where the uncertainty is dominated by that of the Hg measurement. The results are tabulated in Table 2, together with the CIPM recommended frequencies and with a synthetic ratio calculated for the Hg/Yb frequency ratio that we did not yet measure directly.

Table 2. Frequency ratio matrix showing $y = \nu_a/\nu_b$ and the respective fractional uncertainties for $^{87}$Sr, $^{171}$Yb and $^{199}$Hg clocks determined in this work. A synthetic ratio $\nu_{\mathrm{Hg}}/\nu_{\mathrm{Yb}}$ (set in angle parentheses) is calculated from the ratios measured for $\nu_{\mathrm{Hg}}/\nu_{\mathrm{Sr}}$ and $\nu_{\mathrm{Yb}}/\nu_{\mathrm{Sr}}$. $u_c$ is the measurement uncertainty. The CIPM recommended frequencies are also given for reference.

| | | Numerator $\nu_a$ | | |
|---|---|---|---|---|
| | | **$^{87}$Sr** | **$^{171}$Yb** | **$^{199}$Hg** |
| Denominator $\nu_b$ | **$^{87}$Sr** | (1) | 1.207 507 039 343 337 76(29)<br>$u_c/y = 2.4 \times 10^{-16}$ | 2.629 314 209 898 909 60(22)<br>$u_c/y = 8.4 \times 10^{-17}$ |
| | **$^{171}$Yb** | - | (1) | [2.177 473 194 134 565 07(54)]<br>$[u_c/y = 2.5 \times 10^{-16}]$ |
| | **$^{199}$Hg** | - | - | (1) |
| | **CIPM s$^{-1}$** Ref. [59] | 429 228 004 229 873.4(4)<br>$u_c/\nu_{\mathrm{Sr}} = 1.0 \times 10^{-15}$ | 518 295 836 590 865.0(1.4)<br>$u_c/\nu_{\mathrm{Yb}} = 2.7 \times 10^{-15}$ | 1 128 575 290 808 162(19)<br>$u_c/\nu_{\mathrm{Hg}} = 1.7 \times 10^{-14}$ |

## 4. Frequency comparison between two distant Sr clocks via a 30-km-long optical fibre link

According to general relativity, for two clocks with a height difference of $\Delta h$, the higher clock runs faster by $\Delta\nu/\nu_0 = g\Delta h/c^2 \approx 1.1 \times 10^{-16}\mathrm{m}^{-1}\Delta h$, where $\nu_0$ is the clock frequency, $g$ is the gravitational acceleration, and $c$ is the speed of light. Such gravitational red shift was investigated as early as 1960 [60] and its application to geopotential measurements was discussed in 1985 [25]. Applying high-precision optical-clocks, 30-cm height difference has been observed [23] inside a laboratory. Frequency comparison of remote optical clocks has been demonstrated between institutes separated by up to 24 km [24, 61]. Relying on the reproducibility of cryogenic Sr optical lattice clocks demonstrated in Sect. 2, it may be possible to resolve a relative frequency difference of $2 \times 10^{-18}$ for an averaging time of $6000$ s, which corresponds to a height difference of $2$ cm. Networking such clocks allows mapping and monitoring of the temporal variation of gravitational potential, serving as a new tool for relativistic geodesy.

This section describes frequency comparisons between cryogenic Sr clocks operated at RIKEN and the University of Tokyo (UTokyo), separated by 15 km and with a clocks' height difference of about 15 m. In particular, we focus on the high stability of frequency comparisons enabled by optical lattice clocks. While usual optical-clock comparisons [21, 24, 61-63] employ frequency combs that convert clock laser frequencies to a telecommunication frequency at around 200 THz (1.5 μm), we directly linked two Sr clocks at $2\lambda_{\mathrm{Sr}} = 1397$ nm [64], the sub-harmonic of the Sr-clock transition wavelength $\lambda_{\mathrm{Sr}} = 698$ nm, to avoid the stability degradation due to the optical frequency comb. As a result, clocks at both sites are driven by the same clock laser, which allows suppression of the Dick effect by synchronous interrogation.

A schematic of the synchronous frequency comparison between the two distant clocks via the fibre link is shown in Fig. 4. At RIKEN, the transfer laser at 1397 nm, which is phase-locked to a clock laser at 698 nm after frequency doubling (See Fig. 2), is sent to UTokyo, where a repeater laser is heterodyne-locked to the transfer laser and a part of the repeater laser is sent back to RIKEN. At RIKEN, a fibre noise cancellation system [58] uses the beat signal of the transfer and repeater lasers to feedback-control the effective fibre length between the two sites. A propagation delay-time of $\tau_{\mathrm{d}} = 0.15$ ms through the 30-km-long fibre limits the servo bandwidth to be less than $1/(4\tau_{\mathrm{d}}) = 1.7$ kHz. The typical phase noise added to light propagating through the fibre is $S_\phi^{\mathrm{free}}(f) = 6 \times 10^2 (f/\mathrm{Hz})^{-2}\ \mathrm{rad}^2/\mathrm{Hz}$ for Fourier frequencies $f < 100$ Hz, which is suppressed to $S_\phi^{\mathrm{stab}}(f) = 2 \times 10^{-4}\ \mathrm{rad}^2/\mathrm{Hz}$ [64], in good agreement with the theoretical limit [65]. Assuming the phase noise of the clock laser $S_\phi^{\mathrm{laser}}(f) = [0.15\,(f/\mathrm{Hz})^{-3} + 0.02(f/\mathrm{Hz})^{-2}]\ \mathrm{rad}^2/\mathrm{Hz}$, which corresponds to $1 \times 10^{-15}$ frequency stability, we achieve $S_\phi^{\mathrm{stab}}(f) < S_\phi^{\mathrm{laser}}(f)$ for $f < 10$ Hz. This means that the residual fibre noise has little effect on the frequency stability of the Sr clock at UTokyo, as it is mostly the phase noise at $f < T_{\mathrm{i}}^{-1}$ that contributes to the Dick effect [57], where $T_{\mathrm{i}} = 0.3$ s is the interrogation time of the clock transition.

In order to synchronously interrogate the two remote clocks, we send a clock-trigger signal at $1550$ nm, which is superimposed on the transfer laser at $2\lambda_{\mathrm{Sr}} = 1397$ nm using a wavelength division multiplex (WDM) coupler at each end. With this clock-trigger signal, the two distant clocks are synchronized including the propagation delay-time $\tau_{\mathrm{d}}$ of transfer laser. The delay-time difference between 1397 nm and 1550 nm lasers due to fibre dispersion is as small as 0.2 μs, which is negligible to share the low frequency laser noise at $f < T_{\mathrm{i}}^{-1}$.

At both ends, clock laser frequencies are independently stabilized to the centre of the clock transition by using frequency shifters. The laser frequency and excitation fraction of the clock transition are recorded as a time series. The Allan deviation of the frequency difference between the two distant clocks is shown in Fig. 5(a). In synchronous operation, the stability improves as $\sigma_y(\tau) = 1.2 \times 10^{-15}(\tau/\mathrm{s})^{-1/2}$ for an averaging time $\tau$ and reaches $2.2 \times 10^{-17}$ at $\tau = 3000$ s, which is an order of magnitude better than that in asynchronous operation. Figure 5(b) shows the correlation diagram of the excitation fractions of the two synchronized clocks. The correlation of the excitation fractions cancels out the Dick effect, which improves the Allan deviation of frequency comparison in a synchronous operation.

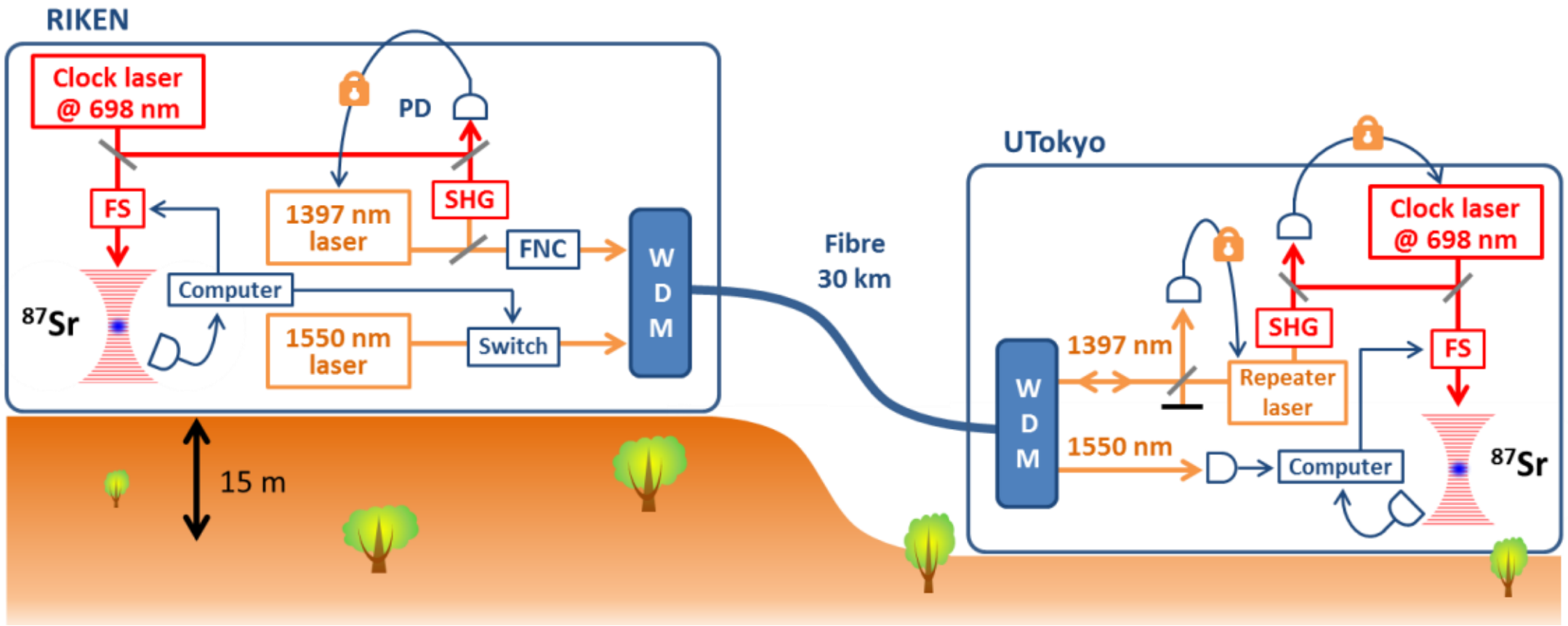

Fig. 4. Schematic of synchronous operation of two distant clocks at RIKEN and UTokyo via a 30-km-long optical fibre link. The laser at 1397 nm is sent to share the clock laser frequency at 698 nm. The laser at 1550 nm is used to send a timing signal for synchronous operation between the distant clocks. FNC, fibre noise canceler; WDM, wavelength division multiplexer.

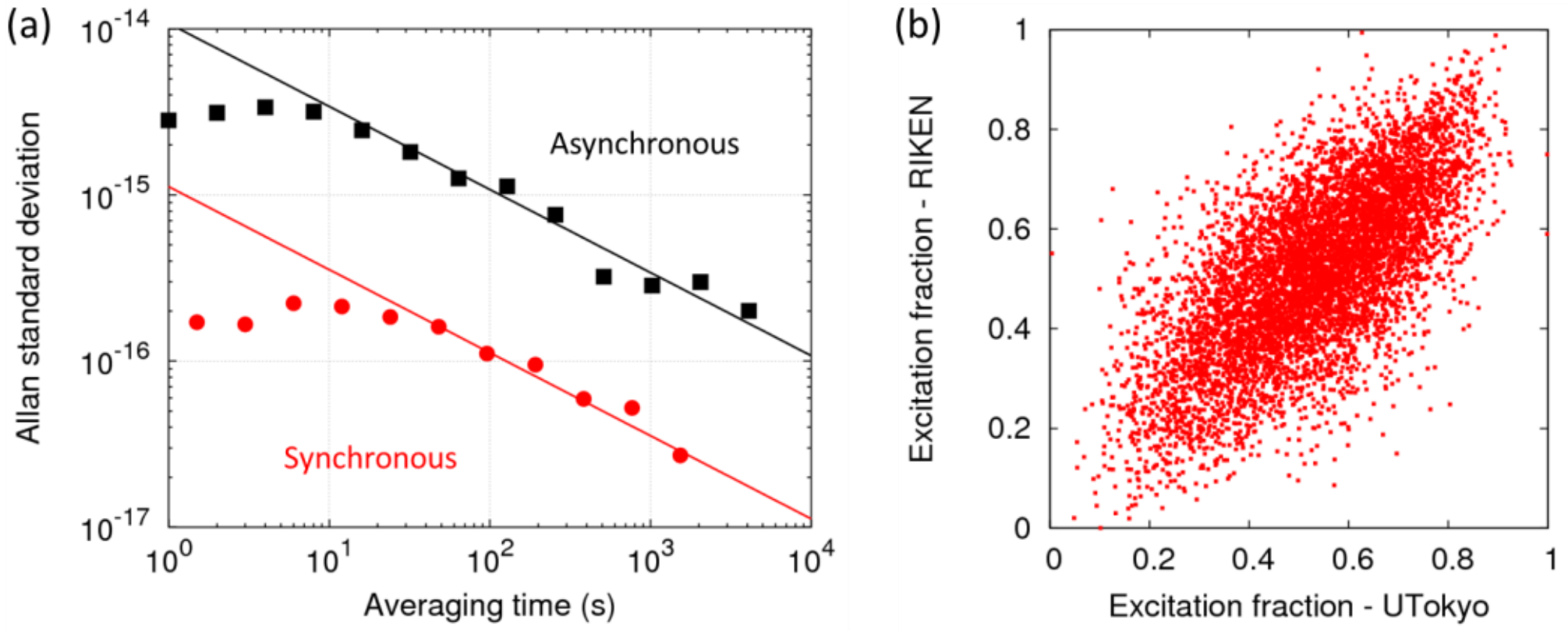

Fig. 5. (a) Allan deviation of the frequency difference between the two clocks as a function of an averaging time in asynchronous (squares) and synchronous (circles) operations. (b) Correlation diagram of the excitation fractions of the two clocks.

## 5. Summary and outlook

We have reviewed current status of optical lattice clocks with neutral Sr, Yb and Hg atoms, and the frequency comparisons between the clocks. Connecting three types of clocks directly by an optical frequency comb, the ratios of the clock frequencies are determined beyond the limitations of the SI second. Gathering the results of such comparisons in a matrix table of clock frequency ratios allows the calculation of optimal synthetic values for frequency ratios that have not yet been measured directly, as well as improved consistency checks. These efforts should be an essential step towards redefinition of the SI second. In addition, the long-term monitoring of frequency ratio will provide an invaluable resource in the search for a temporal variation of the fundamental constants. Further investigation of the systematic effects such as hyperpolarizability and multipolar light shifts [66] and other sources of uncertainties will determine the best atomic species suitable for optical lattice clocks.

Accurate clocks and a frequency link between them are the essential building blocks for relativistic geodesy. For a practical use in monitoring the temporal variation of the gravitational potential, speedy comparison between remote clocks are required, which can be accomplished by synchronous comparison as demonstrated in Sec. 4. In parallel, we are developing technologies to miniaturize optical lattice clocks, aiming at transportable clocks. Such an endeavor will allow investigation of mapping resources, cavities and magma chambers underneath the earth's crust by measuring the gravitational red shift caused by the local change of gravitational potential. As a possible direction towards the clock miniaturization, we have started the development of "fibre clocks" using a hollow-core photonic crystal fibre in which atoms are confined by the magic wavelength lattice [67]. Such transportable ultraprecise atomic clocks will open the way of clocks' application to a gravitational potential meter.

**Acknowledgements**

This work was supported in part by the FIRST Program of the Japan Society for the Promotion of Science and by the Photon Frontier Network Program of the Ministry of Education, Culture, Sports, Science and Technology, Japan.